\documentclass[preprint]{aastex63}
\shorttitle{}
\shortauthors{}
\graphicspath{{./}{figures/}}

\begin{document}

\title{Oxidation of the Interiors of Carbide Exoplanets}

\correspondingauthor{Harrison Allen-Sutter; Sang-Heon Shim}
\email{hallensu@asu.edu, shdshim@gmail.com}

\author[0000-0003-3809-0446]{H. Allen-Sutter}
\affiliation{School of Earth and Space Exploration, Arizona State University, Tempe, AZ, 85287}

\author{E. Garhart}
\affiliation{School of Earth and Space Exploration, Arizona State University, Tempe, AZ, 85287}

\author{K. Leinenweber}
\affiliation{Eyring Materials Center, Arizona State University, Tempe, AZ, 85287}

\author{V. Prakapenka}
\affiliation{GeoSoilEnviroCARS, University of Chicago, Chicago, IL, 60637}

\author{E. Greenberg}
\affiliation{GeoSoilEnviroCARS, University of Chicago, Chicago, IL, 60637}

\author[0000-0001-5203-6038]{S.-H. Shim}
\affiliation{School of Earth and Space Exploration, Arizona State University, Tempe, AZ, 85287}



\begin{abstract}
Astrophysical measurements have shown that some stars have sufficiently high carbon-to-oxygen ratios such that the planets they host would be mainly composed of carbides instead of silicates.
We studied the behavior of silicon carbide in the presence of water under the high pressure--temperature conditions relevant to planetary interiors in the laser-heated diamond-anvil cell (LHDAC). When reacting with water, silicon carbide converts to silica (stishovite) and diamond at pressures up to 50~GPa and temperatures up to 2500~K: 
$\mathrm{SiC} + 2\mathrm{H}_2\mathrm{O} \rightarrow \mathrm{SiO}_2 + \mathrm{C} + 2\mathrm{H}_2$.
Therefore, if water can be incorporated into carbide planets during their formation or through later delivery, they could be oxidized and have mineralogy dominated by silicates and diamond in their interiors. 
The reaction could produce CH$_4$ at shallower depths and H$_2$ at greater depths \replaced{. These}{which} could be degassed from the interior, causing the atmospheres of the converted carbon planets to be rich in reducing gases. Excess water after the reaction can be stored in dense silica polymorphs in the interiors of the converted carbon planets.
\added{Such conversion of mineralogy to diamond and silicates would decrease the density of carbon-rich planet, making the converted planets distinct from silicate planets in mass-radius relations for the 2--8 Earth mass range.}
\end{abstract}

\keywords{carbon planet, silicon carbide, water, silica, diamond, atmosphere}


\section{Introduction} \label{sec:intro}

Carbon-rich planets could exist in extra-solar systems containing either stars with high C/O ratios \citep{bond2010carbonplanets} or proto-planetary discs of Sun-like stars with locally elevated C/O ratios \citep{Kuchner:2005wu}.
\citet{bond2010carbonplanets} suggested a significant population of planet-hosting stars have C/O ratios well over 1;
however, recent studies have called into question the abundance of those stars \citep{fortney2012carbon,nissen2013carbon,teske2014c,suarez2018c,nissen2018}.
\replaced{In any case, systems with $\mathrm{C/O} > 0.8$ certainly exist \citep{young2014astrochem}}{While they seem to comprise no more than 12--17\% of planetary systems 
\citep{wilson2016,stonkute2020}, 
some systems with $\mathrm{C/O} > 0.8$ likely exist \citep{young2014astrochem}. 
Because our solar system does not host carbon-dominated planets, relatively little is known about the type of structure and dynamics that make up the surface and interior of these planets.}

In those carbon-rich planets, silicon carbide (SiC) can be the major mantle phase \citep{larimer1975effect}.
Therefore, high-pressure polymorphs of SiC have been studied extensively at high pressure-temperature $(P{-}T)$  in recent years \citep{nisr2017thermal,daviau2017decomposition,miozzi2018equation,DaviauSiCSiO2,kidokoro2017}
to understand the interiors of carbon-rich planets. 
On the other hand, the oxidation of SiC under hydrothermal conditions has been known in the materials science literature for many decades.  
For example, SiC oxidizes in the presence of water at temperatures as low as 700\,K and pressures as low as 0.01\,GPa to form silica and gasses  \citep{yoshimura1986oxidation}. 
\added{The well-known reaction at very low pressure conditions raises an important question on the stability of SiC in the interior of carbon-rich planets when water exists or is delivered.
Recent studies have found that SiC could react with oxides in carbon rich bulk compositions and form graphite at high temperature and under 2~GPa \citep{hakim2018capturing,hakim2019mineralogy}. 
However, these studies were conducted at low pressures and did not consider the effect of water.}
Therefore, it is important to further investigate if SiC would remain the main phase at the high $P{-}T$ conditions of the interiors of carbide planets in the presence of water.
Here, we report experimental investigation on SiC + H$_2$O mixtures at high $P{-}T$ in the laser-heated diamond-anvil cell (LHDAC) combined with synchrotron X-ray diffraction (XRD) and micro-Raman spectroscopy. 

\section{Materials and Methods} \label{sec:meth}

\begin{table}[!tb]
\caption{Experimental runs performed in this study.
The temperatures of the LHDAC runs were obtained from the gray-body radiation from the samples except for the DAC13 run.
For DAC13 we estimated from the intensity of thermal radiation.
The estimated uncertainties for the temperatures are 100--150~K\@.
The estimated uncertainty for pressure is approximately 2--5~GPa.
S.M.: starting material, XRD: X-ray diffraction, Raman: micro-Raman spectroscopy, $T$: temperature, $P$: pressure or pressure range, $t$: time duration of heating, and $P$ scale: pressure scale calibrant.  }\label{runtable}
\begin{tabular}{l c c c c c c c} 
\\ \toprule
Run   & S.M. & $t$ (min) & $T$ (K) & $P$ (GPa)   & $P$ scale & Analysis                \\ \hline 
DAC2  & SiC-6H & 10             & 1050--1600 & 34.5--44.5 & Au          & XRD              \\  
DAC3  & SiC-6H & 10             & 1450--1950 & 39--43.5   & Au          & XRD              \\  
DAC4  & SiC-6H & 6              & 1550--1850 & 41.5--48   & Au          & XRD              \\  
DAC5  & SiC-6H & 7              & 1450--2100 & 42--49     & Au          & XRD              \\  
DAC6  & SiC-6H & 9              & 1575--1975 & 41--47     & Au          & XRD              \\  
DAC7  & SiC-6H & 4              & 1700--2675 & 43--47     & Au          & XRD              \\  
DAC8  & SiC-6H & 9              & 1525--2200 & 42--47     & Au          & XRD              \\  
DAC9  & SiC-6H & 9              & 1125--1600 & 28         & Au          & XRD              \\  
DAC10 & SiC-6H & 7              & 1075--1450 & 26         & Au          & XRD              \\  
DAC11 & SiC-6H & 7              & 1150--1600 & 24.5--28.5 & Au          & XRD              \\  
DAC12 & SiC-6H & 9              & 1125--1625 & 26         & Au          & XRD              \\  
DAC13 & SiC-6H & 7              & 1350--1650 & 28--29     & Au          & XRD              \\  
DAC14 & SiC-6H & 20             & 1325--1600 & 38         & Ruby        & Raman            \\  
DAC15 & SiC-6H & 20             & 1275--1675 & 38         & Ruby        & Raman            \\  
DAC16 & SiC-6H & 20             & 1350--1550 & 38         & Ruby        & Raman            \\  
DAC17 & SiC-6H & 21             & 1350-1550  & 38         & Ruby        & Raman            \\  
DAC18 & SiC-3C & 15             & 1400       & 20         & Au          & XRD              \\  
\hline 
\end{tabular}
\end{table}

Starting materials were pure synthetic SiC (Alfa, purity 99.8$\%$) of the hexagonal $\alpha$ phase (SiC-6H) or cubic $\beta$ phase (SiC-3C).
For LHDAC experiments, the SiC powder was mixed with gold powder (10~wt\%) as a laser coupler and pressure calibrant. 
The SiC + gold powder mixture was cold-pressed into foils with approximately 10\,$\mu$m of thickness. 
The foils were loaded into 125\,$\mu$m and 260\,$\mu$m holes drilled in a rhenium gasket which had been indented by diamond anvils with 200\,$\mu$m and 400\,$\mu$m diameter culets, respectively. 
The holes were then filled with deionized water. 
Samples were compressed to pressures between 20 and 40\,GPa at 300~K before laser heating. 
A total of 18 LHDAC runs were performed (Table~\ref{runtable}).

XRD patterns were collected at high $P{-}T$ in double-sided laser-heated DAC (diamond-anvil cell) at the 13-IDD of the GeoSoilEnviroConsortium for Advanced Radiation Sources (GSECARS) sector at the Advanced Photon Source (APS).
Monochromatic X-ray beams of wavelength 0.4133\,{\AA}\ or 0.3344\,{\AA}\ were focused on the sample in LHDAC\@. 
Near-infrared laser beams were coaxially aligned and focused with the X-ray beams for in situ laser heating. 
Temperatures were estimated by fitting thermal spectra from both sides to the gray-body equation \citep{prakapenka2008advanced}. 2-D diffraction images, collected from a Dectris Pilatus detector, were integrated into 1-D diffraction patterns using DIOPTAS \citep{Prescher2015DIOPTAS}. 
Using the CeO$_{2}$ and LaB$_{6}$ standards, we corrected for tilt of the detector and sample-to-detector distance. 
The diffraction peaks were fitted with a pseudo-Voigt profile function to determine the unit-cell parameters in PeakPo \citep{PeakPo}. 
The unit-cell parameter fitting was conducted based on the statistical approaches presented in \citet{holland1997unit}. 
Pressure was calculated by combining the measured unit-cell volume of gold with its equation of state \citep{ye2017intercomparison} using Pytheos \citep{pytheos}. 
In some DAC runs, pressure was estimated from ruby spectra at 300~K \citep{piermarini1975ultrahigh}.

Micro-Raman measurements were conducted for the phase identification of the recovered samples from DAC runs 14--17 at Arizona State University (ASU\@). 
We used a solid-state (frequency doubled Nd:YAG) laser with a 532\,nm monochromatic beam, set to a laser power of 50--100~mW (5--10~mW at the sample), as an excitation source. 
Measurements were conducted using an 1800\,grooves/mm grating. 
The spectrometer was calibrated using the neon emission spectra. 
We calibrated pixel-to-pixel sensitivity differences in the CCD detector using the spectrum of a glass with well-known fluorescence intensities at different wavenumbers. 
Spectra were measured at different wavenumber ranges: 100--1000~cm$^{-1}$ for SiC and SiO$_2$, 1000--1500~cm$^{-1}$ for diamond, and 2000--4000~cm$^{-1}$ for H$_2$O, CH$_4$, and H$_2$. The typical acquisition time was 50--100~seconds.

\added{We calculated the mass-radius relations of planets composed of relevant materials.
In the models, planets are composed of a single homogeneous layer. 
The equations-of-state parameters used for these calculations were from: \citet{dewaele2008high} for diamond, \citet{miozzi2018equation} for silicon carbide in B1 structure, \citet{dewaele2006quasihydrostatic} for iron metal, \citet{stixrude2011thermodynamics} for silica (seifertite) and MgSiO$_3$ (bridgmanite), and \citet{hemley1987static} for ice. 
We chose the B1 phase of SiC and seifertite because they would be the dominant phases in 2--8 Earth mass planets  \citep{daviau2017zinc,miozzi2018equation,kidokoro2017,grocholski2013stability}.
The equations of state were calculated using the Burnman package \citep{burnman}. 
We did not include thermal effects, because internal temperatures of the exoplanets are not well known and the thermal effects are much smaller than pressure effects on density of minerals for the mass range we consider.}

\begin{figure}[!ht]
\centering
\includegraphics[width=0.9\textwidth]{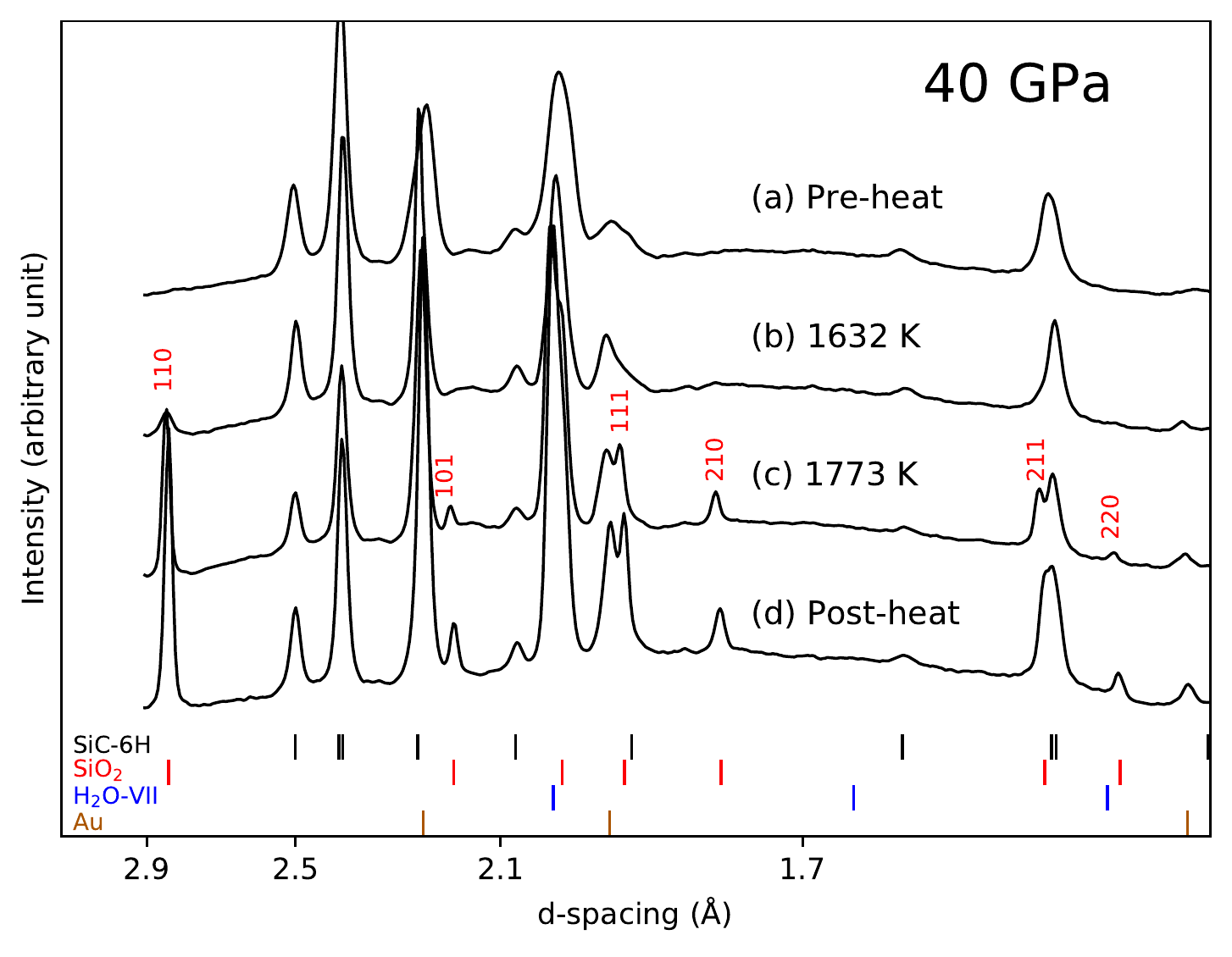}
\caption{X-ray powder diffraction patterns measured at in-situ high pressures and high temperatures: (a) the starting material before heating, (b) the sample just after heating began, (c) the sample 10~min later, and (d) the sample after heating. 
The energy of X-ray beam was 37~keV. 
The colored vertical bars show the expected diffraction peak positions of phases.
The Miller indices of main stishovite lines are shown to highlight the appearance of those lines during heating.}
\label{XRD}
\end{figure}

\section{Results} \label{sec:results}

X-ray diffraction (XRD) patterns showed the conversion of SiC into SiO$_2$ stishovite from every run across our entire $P{-}T$ range regardless of the polymorphs of SiC. For example, at 40~GPa before heating, the only peaks observed were from SiC-6H and H$_2$O ice VII (starting mixture) together with Au (pressure standard) (Figure\,\ref{XRD}). 
As soon as the heating began, the $110_\mathrm{stv}$ diffraction peak at 2.8~{\AA} was immediately visible (the numbers are the Miller index and the subscript notes the phase).
The $110_\mathrm{stv}$ line is diagnostic of stishovite in our diffraction patterns for the identification of the phase because it does not overlap with lines from any other phases and it is the most intense peak for stishovite.
After about 5~minutes of heating, other stishovite diffraction lines\textemdash such as the 101, 111, 210, 211, and 220 peaks\textemdash were all visible. 
All SiO$_2$ peaks continued to grow as heating continued, and they persisted after quench to room temperature at high pressure. 

\begin{figure}[!htbp]
\centering
\includegraphics[width=1\textwidth]{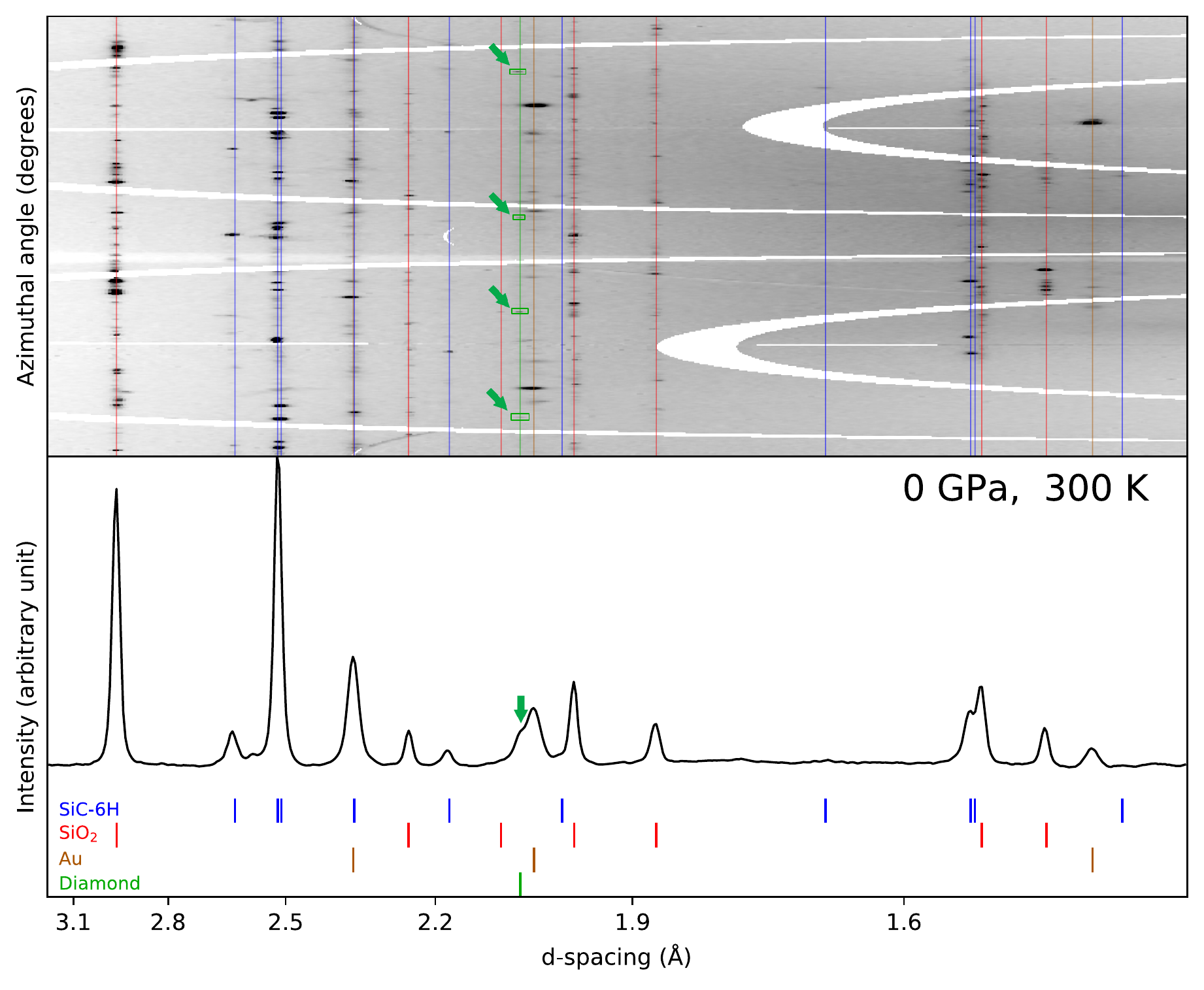}
\caption{Diffraction pattern (bottom) of a sample heated to 1800~K at 40~GPa then recovered to room temperature at 1~bar. 
Diamond diffraction spots can be seen highlighted by the green rectangles and arrows in the unrolled 2-D diffraction image (top). These spots were only present in heated regions of the sample.
The colored bars in the 1-D integrated pattern and the vertical lines in the 2-D unrolled image show the expected peak positions of phases. \deleted{Stishovite in the recovered sample has unit-cell parameters of $a=4.1829(2)$~{\AA} and $c=2.6659(3)$~{\AA}, and a unit-cell volume of $46.646(8)$~{\AA$^3$} according to the fitting.}
}
\label{DXRD}
\end{figure}

\added{In majority of our runs, temperatures were higher than the melting of H$_2$O ice.}
In \replaced{some LHDAC runs}{run DAC2}, \replaced{we increased temperature above}{temperatures were lower than} the melting temperature of H$_2$O ice reported in \citet{h20melt}. 
\deleted{The melting could be also inferred from a plateau in laser power and temperature relations as previous studies have found \citep{walter2015}.}
\replaced{Below the melting temperature of H$_2$O ice}{In this run}, stishovite lines still appeared.
Therefore, the reaction occurs in both solid and liquid regimes of H$_2$O; 
\added{however, there is uncertainty in the melting curve of water. 
If melting temperature of water is low at high pressure as some other studies suggested \citep{goncharov2005dynamic,lin2005melting}, all of our temperatures would fall in the liquid water regime.}

In all the LHDAC experiments, the stishovite peaks continued to grow until the end of the run, which we limited to 30~min for the mechanical stability of the DAC\@.
The continuous growth of the peak intensity indicates that stishovite is stable over SiC in the presence of H$_2$O. 
\added{The continued presence of SiC is likely related to kinetics due to the short heating duration.} 
We expect that SiC would convert completely to stishovite with sufficiently long heating. 

After heating to 1800~K at 42~GPa, we decompressed the sample to 1~bar and measured XRD patters of the recovered sample (run DAC4). \replaced{
The DAC was opened to remove liquid water through evaporation.
We then close the sample chamber again (but still 1~bar) for XRD measurements of the recovered samples.}{We opened the DAC to remove liquid water through evaportion, and then closed the sample chamber again (but still at 1 Bar) for XRD measurements of the recovered samples.} 
All the phases observed at high pressure remained at 1~bar, including stishovite. 

\added{The lattice constants of the recovered samples were obtained by fitting the XRD patterns: $a=4.1829(2)$~{\AA}, $c=2.6659(3)$~{\AA}, and $V = 46.646(8)$~{\AA$^3$} in stishovite and $a=3.081(8)$~{\AA}, $c=15.127(7)$~{\AA}, and $V = 124.4(3)$~{\AA$^3$} in SiC ($a$ and $c$ are axial lengths and $V$ is unit-cell volume).}
In those patterns, we also found some diffraction spots that indexed well with the diamond 111 line (Figure~\ref{DXRD}).
This observation indicates that diamond exists as a few small single crystal grains likely grown in an H$_2$O medium.
The line can also be clearly identified in the integrated 1D diffraction patterns. 
At in-situ high pressure, it was difficult to unambiguously identify the diffraction lines of diamond formed through the reaction, because the most intense diffraction peak of diamond exists at the same $2\theta$ angle range as the main diffraction line of H$_2$O ice VII. 

\added{Although phase fractions can be calculated from the diffraction intensities from randomly oriented crystallites in powder, it was difficult to apply the method for our diffraction patterns.
As explained above, the diamond diffraction is from a small number of single crystals. 
In this case, the diffraction intensity can be highly sensitive to the preferred orientations of these grains and therefore cannot be used for estimating the phase fraction.
The intensity of the diamond peak can be affected by other factors. 
Low-pressure experiments on this system showed the production of methane \citep{yoshimura1986oxidation}.
\cite{hirai2009polymerization} showed that the conversion of methane to diamond requires heating to elevated temperatures at high pressures. 
If the relatively low temperature in this work could not have supplied enough heat to overcome the slow kinetics of the transformation of CH$_4$ to diamond, some C may exist in fluid CH$_4$ which could be convected away in the liquid H$_2$O medium.}  

\begin{figure}[bhtp!]
\centering
\includegraphics[width=0.5\textwidth]{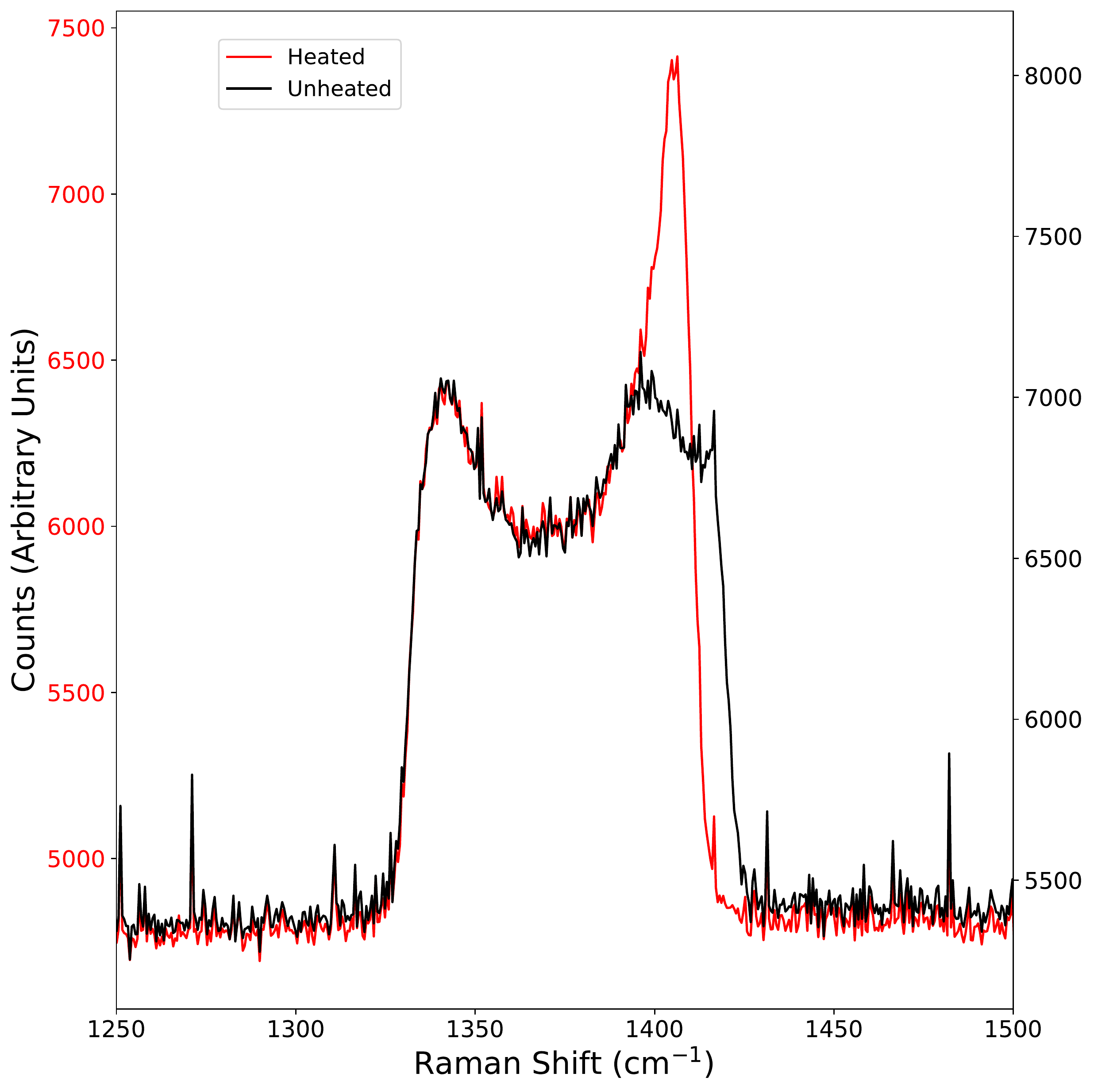}
\caption{High-pressure Raman spectra of the sample synthesized at 40~GPa and 1500~K. The black spectrum was measured at an unheated portion of the sample and the red spectrum was measured at a heated portion.  
The sharp peak at 1400~cm$^{-1}$ in the red spectrum is from  pressurized diamond crystals formed from reaction~\ref{eq:reaction}. 
}\label{DRaman}
\end{figure}

Micro-Raman spectroscopy of the recovered samples confirmed the presence of stishovite. 
In runs DAC 14--18 (Table\,\ref{runtable}), we conducted Raman measurements at in-situ high pressure at 40~GPa in a DAC after laser heating to 1500~K for 20~min (Figure~\ref{DRaman}). 
In order to reduce the Raman scattering from thick diamond anvils, we used a confocal setup. The black spectrum in Figure~\ref{DRaman} was measured at an unheated area of the sample. The observed broad feature is from parts of diamond anvils with different stress conditions along pressure gradients.
The depth resolution achieved in the confocal setup is approximately 30~$\mu$m, which is still somewhat greater than the thickness of the sample in DAC, 5--10~$\mu$m.
Therefore, even if the focal plane is set on the surface of the sample in DAC, some intensity from the tip of the diamond anvils is expected to be detected.
Indeed, the small increase in intensities near the highest wavenumber ($\sim$1400~cm$^{-1}$) should be from the tip of the diamond anvil which is under the highest stress.
In a heated area, we observed a much more pronounced phonon peak intensity at $\sim$1400~cm$^{-1}$. This suggests that a majority of the intensity should come from the compressed diamond crystals in the sample chamber \citep{boppart1985raman} formed from the SiC + H$_2$O reaction, rather than from the diamond anvil.

Both the XRD and Raman observations reported above indicate reaction between SiC and H$_2$O:
\begin{equation}\label{eq:reaction}
    \mathrm{SiC} + 2\mathrm{H}_2\mathrm{O} \rightarrow \mathrm{SiO}_2\  (\mathrm{stishovite}) + \mathrm{C}\ (\mathrm{diamond}) + 2\mathrm{H}_2.
\end{equation}
The reaction also predicts the formation of hydrogen.
While we observed SiO$_2$ stishovite and diamond, we did not directly observe hydrogen.
Hydrogen is difficult to detect in XRD because of its extremely small scattering cross section compared with other materials in the sample chamber of our experiments.
Molecular hydrogen has a Raman mode at 4100--4300~cm$^{-1}$ at the pressure range of this study \citep{goncharov2001H2,gregoryanz2003raman}. However, we could not find the hydrogen peak in our Raman measurements.
Hydrogen likely diffused out from the heated spot to the water medium and therefore can be diluted to a smaller fraction at any given spot. In this case, the hydrogen peak would be very difficult to detect.

\citet{spektor2011ultrahydrous} reported that stishovite can be hydrated and store up to 1.3~wt\% H$_2$O in the crystal structure. 
Even greater H$_2$O storage capacities of dense silica polymophs were reported in recent LHDAC experiments, up to 8--13~wt\% \citep{NISR2020}.
Studies have shown that such significant hydration can expand the unit-cell volume of stishovite and alter the axial ratio $(c/a)$ \citep{NisrHstv,spektor2016formation,NISR2020}. 
The diffraction patterns of our recovered LHDAC samples at 1~bar showed unit-cell volumes larger than those reported for anhydrous stishovite \citep{andrault2003equation,grocholski2013stability}. 
For example, stishovite in the recovered sample from run DAC6, heated to 1800~K at 40~GPa, was expanded by 0.28\% compared to the anhydrous unit-cell volume \citep{andrault2003equation}. 
Based on the relationship between the unit-cell volume expansion and H$_2$O content reported by \citet{NisrHstv}, we obtained  0.5--0.6~wt\% H$_2$O in the stishovite phase.
Therefore, the stishovite formed from the SiC + H$_2$O reaction contains some amount of H$_2$O in the crystal structure.

\section{Discussion} \label{sec:discussion}

\begin{figure}[!htbp]
\centering
\includegraphics[width=1\textwidth]{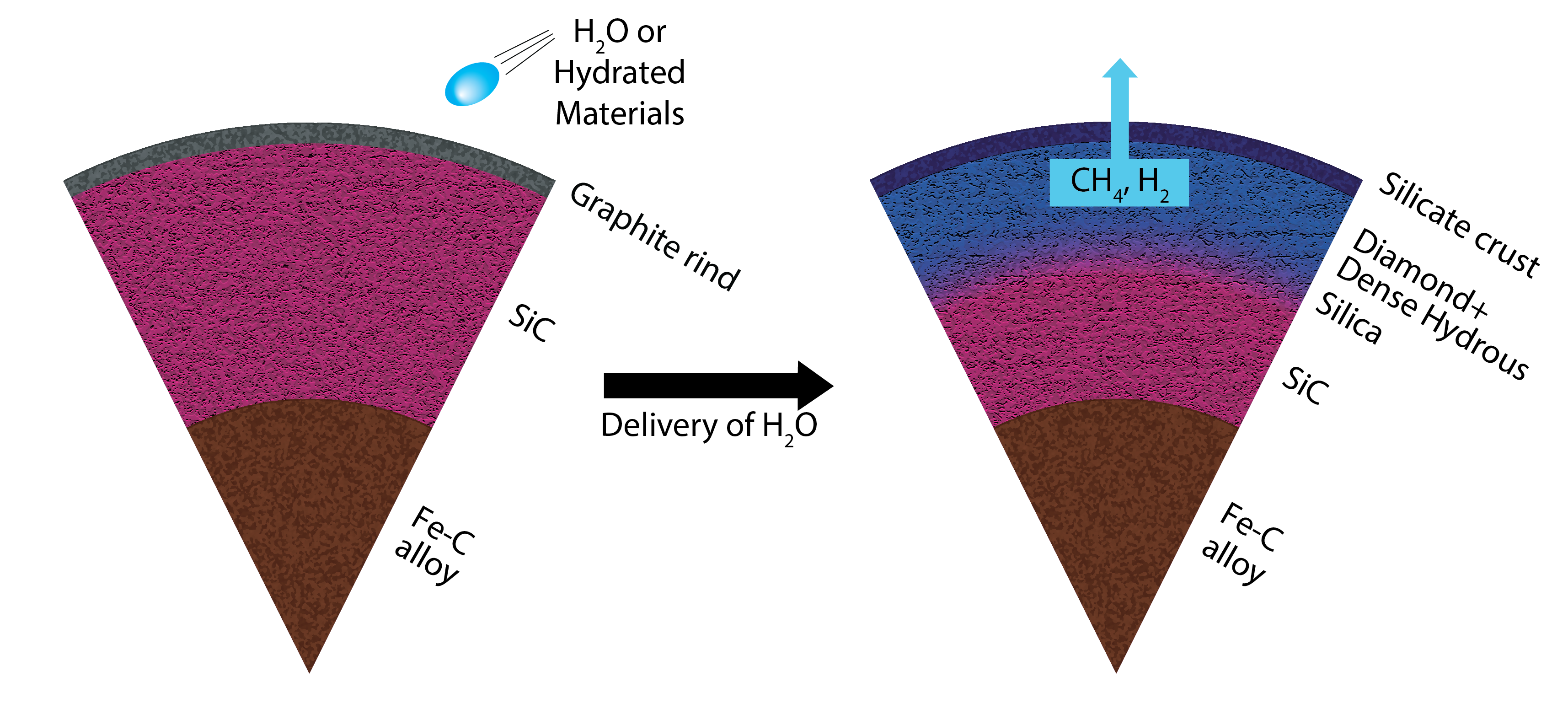}
\caption{An example of a carbon planet with SiC as the major mantle phase (left). 
After bombardment with water-rich materials, the upper portion of the mantle, which was exposed to water, transforms from carbide to silicate and diamond (right). The reaction will also produce CH$_4$ at shallower depths and H$_2$ at greater depths.
The reducing volatiles may be degassed from the interior and incorporated into the atmosphere.
The dense silica polymorphs in the mantle could then store a large amount of H$_2$O in their crystal structures.
}
\label{planets}
\end{figure}



\replaced{A significant}{Some} population (12--17\%) of stars \added{may} have C/O ratios greater than \replaced{1}{0.8} \citep{bond2010carbonplanets,Petigura_2011,wilson2016,stonkute2020}, and the mineralogy of planets hosted around these stars would be dominated by carbides \citep{Kuchner:2005wu,bond2010carbonplanets,wasp-12b,fortney2012carbon,Petigura_2011,duffy2015min}. Therefore, a planet formed under these conditions could have an exotic internal structure and dynamics compared with the planet types observed in the solar system. 
\citet{graphite} suggests that instead of a rocky crust, carbon-rich planets would form a graphite rind as shown in Figure~\ref{planets}. 
This rind could react with hydrogen or water to form a methane-rich atmosphere. 
In the mantles of carbon-rich planets, SiC would be the major phase \citep{larimer1975effect}. 
The core would likely incorporate \replaced{a large amount of}{some} carbon
as well due to \replaced{its abundance and increasing solubility in iron at high pressure}{the abundance of carbon in the system and its solubility in iron at high pressure-temperature}

While there is an inverse relationship between C/O ratios (and consequently carbide abundance) and water abundance \citep{wasp-12b,pekmezci2019statistical}, carbides and water could exist in the same planetary system in significant quantities depending on the C/O ratio, redox conditions, and proportion of available carbon in the solid phase as discussed in \citet{pekmezci2019statistical}. 
In addition, carbide planets can form at a zone with locally elevated C/O ratios relative to the host star due to inherent disk inhomogeneities \citep{Kuchner:2005wu,bond2010carbonplanets}. 
\replaced{In this case, the system could still contain a significant amount of water.}{In this case, the C/O ratio of the system would be lower like our solar system and therefore could contain a significant amount of water.}

Our experiments show that water can react with SiC and convert it to silica + diamond at high $P{-}T$.
Because a similar conversion of SiC by water to silica has also been reported at very low pressure \citep{yoshimura1986oxidation}, the oxidation reaction can likely occur from the shallow depths of the carbide planets.
Two cases can be considered for reaction~\ref{eq:reaction} at a planetary scale: either existence of water during the formation of carbide planets or the delivery of water rich materials at later stages of carbide planet formation, such as late veneer discussed for the Earth \citep{dreibus1987,morbidelli2000,wang2013}.

If water is delivered to carbide planets, the impact will produce high pressure and high temperature locally and induce the reaction.
In regions of the mantle where water reaches SiC, the reaction shown in eq. \ref{eq:reaction} would produce diamond and silica. 
In this case, a carbide planet would experience a chemical change from the outside in.
This process could cause the surface to be covered with silica, while at sufficiently greater depths diamond and silica would exist together as shown in Figure~\ref{planets}. 
Diamond and stishovite have high viscosity and diamond has extremely high thermal conductivity \citep{weidner1982stishovite,mcskimin1972elastic}.
Because of the physical properties, it is unlikely that the diamond + silica-rich mantle would have vigorous convection. 
In the converted planet, the secular cooling would \added{likely} be dominated by conduction, differing from Earth-like planets \citep{Unterborn_2014,nisr2017thermal}. 
\added{Water can alter the rheology of mineral phases \citep{hirth1996water}.
Therefore, if the silica phase contains a large amount of water in the converted planets \citep{nisr2017phase,NISR2020}, it could have a different rheology from anhydrous case.
Therefore, it is important to address the impact of water on silica's rheology in future studies.}

The hydrogen formed in reaction~\ref{eq:reaction} at high pressure would be degassed from the interior and incorporated in the atmosphere.
At pressures below the stability of diamond, the reaction likely produces methane as shown by \citet{yoshimura1986oxidation}: 
\begin{equation}\label{eq:methane}
\mathrm{SiC} + 2\mathrm{H}_2\mathrm{O} \rightarrow \mathrm{SiO}_2 + \mathrm{CH}_4.
\end{equation}
This reaction leads us to believe that at shallower depths and lower temperatures, methane may be produced in hydrated carbon-rich planets. 
As temperature and pressure increase with depth, it is possible that methane can polymerize to form ethane and higher-order hydrocarbons \citep{hirai2009polymerization}. 
Therefore, it is feasible that depending on the depths of the chemical alteration by water, the interior of a carbide planet can produce different reducing volatiles (such as methane and hydrogen).
If they are degassed and incorporated into the atmosphere, the converted carbon-rich planets would have an even more reduced atmosphere. 

\begin{figure}[!htbp]
\centering
\includegraphics[width=1\textwidth]{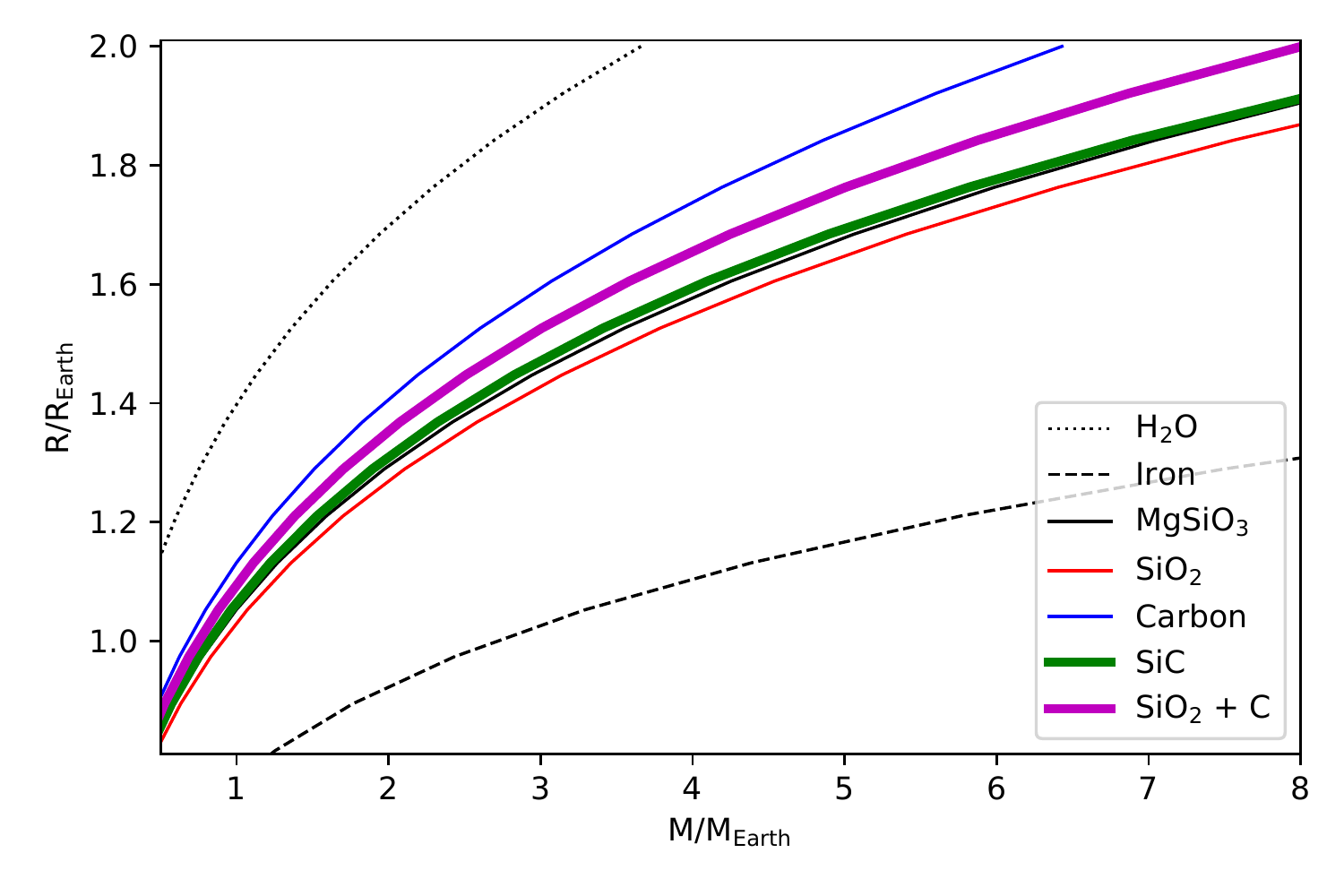}
\caption{Mass-radius relations of planets composed of different materials. 
A planet composed of SiC (the thick green curve) is indistinguishable from one composed of MgSiO$_3$ (the thin black curve) because of the stability of the dense B1 phase at high pressure. 
However, a planet composed of diamond + silica (SiO$_2$ + C; the thick purple curve) would be less dense.
}
\label{mass-radius}
\end{figure}

It is important to note that stishovite can store a large amount of H$_2$O in the crystal structure at high pressure.
A recent study showed that the solubility in dense silica polymorphs increases with pressure, at least up to 100~GPa, and reaches 8--13~wt\% H$_2$O in silica \citep{NISR2020}.
Therefore, a once-carbide planet that has undergone the conversion from SiC to silicate + diamond could store a large amount of water in its mantle. 

\added{An important consideration is how different mineralogy would affect astrophysical observables such as mass, radius, and atmosphere. 
While the atmospheric composition is already reducing and depleted in water in carbide planets \citep{madhusudhan2011high}, the chemical reaction presented here would likely push it further in that direction by sequestering oxygen in the mantle and producing carbon- and hydrogen-rich volatile species in the atmosphere during conversion to silicate + diamond planets. 
As shown in Figure~\ref{mass-radius}, a planet composed of SiC would be indistinguishable from a planet composed of magnesium silicates (MgSiO$_3$). 
The main reason for the similarity is that the dense B1 phase of SiC becomes stable and therefore dominant in the 2--8 Earth mass planets.
However, if a carbide planet undergoes alteration by H$_2$O, the mineralogy would change to SiO$_2$ + diamond while the bulk chemical composition remains the same.
Because of highly incompressible diamond, the converted carbon-rich planet with diamond and silica would become significantly less dense as shown in Figure~\ref{mass-radius}.
The difference may not be sufficiently large compared with the level of uncertainties in the existing data for the mass and radius of exoplanets and the degeneracy in the mass--radius parameteral space but the difference in the mass-radius relations combined with the predicted interior--atmosphere relations for the converted carbon-rich planets provide improved constraints for the future investigations for carbon-rich worlds.}

Silicon carbide would be the main constituent of carbide planets.
However, other elements may exist in the planet.
For example, Mg carbides could become important as the Mg/Si ratio increases.
Some Mg carbides are known to react with water and form oxides and hydroxides at low pressures \citep{rueggeberg1943carbides,lauren1968novel,hajek1980hydrolysis}, similar to the case for Si carbide.
Therefore, it is possible that Mg-rich carbides convert to Mg-oxides and diamond, contributing to the conversion of carbide planets.
If a carbide planet is large enough to exceed \replaced{100}{52--75}~GPa in the mantle, SiC will undergo a phase transition \citep{SiCphasetrans,sekine1997shock,daviau2017zinc,miozzi2018equation,kidokoro2017}.
\replaced{Although we did not consider any polymorphic phase transitions in this work, they}{However, the polymorphic phase transition} would likely not make a significant impact on our implications \added{on chemical reactions in the carbon-rich planets,} due to the outward-in nature of the transformation (Figure~\ref{planets}). 
Future works on high-pressure polymorphs of SiC with water would address the question of how deep the reaction presented here can occur in carbon-rich planets.

\section{Conclusion} \label{sec:conclusion}

Combined with the existing experiments at low pressures, our new experiments at high pressures show that water can convert silicon carbide to silica and diamond.
With our finding that carbide planets will readily convert to silicate planets in the presence of water, the number of carbide planets in existence may be even lower than current predictions. 
Furthermore, a carbide planet could convert to a type of planet which to our knowledge has never been considered before: a planet rich in both diamond and silicates.
The unique mineralogy of the converted carbon-rich planets would make the planets un-Earth-like. 
For example, the mantle of the converted planets would be much more viscous than the Earth-like silicate mantle, because of the physical properties of silica and diamond. 
Because diamond is a main mineral in those these converted, the secular cooling of the planets could be dominated by conduction from high thermal conductivity of diamond.
The atmosphere of the converted planets could be very reducing from the methane and hydrogen degassed from the hydration of the interiors.  
In contrast, a significant amount of water could remain and be stored in the deep mantle of the converted planet because of the large water storage capacity of dense silica polymorphs at high pressures.
\added{During the conversion of a carbon-rich planet, the mineralogy change from carbide to silica + diamond would reduce the density of the planet, because of highly incompressible diamond.
Such a conversion would not change the bulk chemical composition significantly.
Instead, the mineralogy change alone can shift in the mass-radius relations substantially for the carbon-rich world.}

\acknowledgments

We thank J. Dolinsch and J. Tappan for their assistance with high pressure experiments at Arizona State University (ASU). 
We also thank two anonymous reviewers.
This work is supported by NASA Exoplanet Program 80NSSC18K0353. 
The results reported herein benefit from collaborations and information exchange within Nexus for Exoplanet System Science (NExSS) research coordination network sponsored by NASA's Science Mission Directorate. 
Portions of this work were performed at GeoSoilEnviroCARS (The University of Chicago, Sector 13), Advanced Photon Source (APS), Argonne National Laboratory. 
GeoSoilEnviroCARS is supported by the National Science Foundation - Earth Sciences (EAR-1634415) and Department of Energy (DOE) - GeoSciences (DE-FG02-94ER14466). 
This research used resources of the Advanced Photon Source, a U.S. DOE Office of Science User Facility operated for the DOE Office of Science by Argonne National Laboratory under Contract No. DE-AC02-06CH11357. 
We acknowledge the use of facilities within the Eyring Materials Center at ASU\@. 
The experimental data for this paper are available by contacting hallensu@asu.edu.

%

\vspace{5mm}






\bibliography{bib}
\bibliographystyle{aasjournal}




\end{document}